\documentclass{statsoc}

\usepackage[T1]{fontenc}
\usepackage[a4paper]{geometry}
\usepackage{graphicx}
\usepackage[textwidth=8em,textsize=small]{todonotes}
\usepackage{adjustbox}
\usepackage[utf8]{inputenc}
\usepackage{amsmath}
\usepackage{natbib}
\usepackage{enumitem}
\usepackage{array}
\usepackage{setspace}
\usepackage{placeins}
\usepackage{multirow}
\usepackage{tabularray}
\usepackage{wrapfig}
\usepackage{caption}
\usepackage{booktabs}
\usepackage{float}
\usepackage{url}       
\usepackage{fancyhdr} 
\usepackage{hyperref}  
\title[Question Framing and Automatic Coding]{The Impact of Question Framing on the Performance of Automatic Occupation Coding}

\author[O. Kononykhina et al.]{%
\large
Olga Kononykhina$^{1,2}$\thanks{Corresponding author. Mailing Address: Social Data Science and AI Lab (SODA), Ludwigstr. 33, 80539 München, Germany Email: olga.kononykhina@lmu.de}, 
Frauke Kreuter$^{1,2,3}$,
Malte Schierholz$^{1,2}$ 
}

\address{%
\footnotesize
$^1$Ludwig-Maximilians-Universität München, Germany \\
$^2$Munich Center for Machine Learning, Germany \\
$^3$University of Maryland, College Park, USA
}

\begin{document}

\begin{abstract}
Occupational data play a vital role in research, official statistics, and policymaking, yet their collection and accurate classification remain a challenge. This study investigates the effects of occupational question wording on data variability and the performance of automatic coding tools. We conducted and replicated a split-ballot survey experiment in Germany using two common occupational question formats: one focusing on “job title” (\textit{Berufsbezeichnung}) and another on “\textit{berufliche Tätigkeit}” (loosely translated as occupation or occupational task). Our analysis reveals that automatic coding tools, such as CASCOT and OccuCoDe, exhibit sensitivity to the form and origin of the data. Specifically, these tools were more efficient when coding responses to the job title question format than the occupational task format, suggesting a potential way to improve the respective questions for many German surveys. In a subsequent “detailed tasks and duties” question, providing a guiding example prompted respondents to give longer answers without broadening the range of unique words they used. These findings highlight the importance of harmonising survey questions and and ensuring that automatic coding tools are robust to differences in question wording. Further research is needed to optimise question design and coding tools for greater accuracy and applicability in occupational data collection.

\end{abstract}

\section{Introduction}

Work is a fundamental aspect of human life and a key driver of the global economy. Despite its centrality, the collection and aggregation of occupational information for use in official statistics, research, and policymaking remain challenging. While asking someone, ``What do you do?'' might seem straightforward in everyday conversations, responses such as ``I am a teacher'' lack the precision needed for statistical analysis. To inform employment trends, occupational health or migration patterns, occupational data must be collected in a way that ensures both clarity and consistency. Moreover, this information needs to be accurately coded into nationally or internationally recognized classification systems, such as the International Standard Classification of Occupations (ISCO).

In survey methodology, the phrasing of occupational questions affects the clarity and detail of respondents' answers. Accurate coding requires respondents to effectively translate their job familiarity into sufficiently specific descriptions. Occupational questions, commonly structured in two parts—an initial open-ended summary followed by questions about primary tasks—, have been central to censuses and surveys for decades \citep{Tijdens2014}. Research has shown that the phrasing of these questions significantly influences the length and specificity of the responses \citep{Martinez2017}, which in turn affects the accuracy and reliability of occupational coding \citep{conrad2016classifying}.

Traditionally, occupational responses have been manually coded by trained professionals. Although accurate, it is labour-intensive, time-consuming, and prone to human error and variability between coders \citep{Massing2019}. Recent advances in automatic coding methods, from simple text-matching algorithms to machine learning and large language models, aim to address these issues by improving efficiency \citep{schierholz2021machine}.  Despite their promise, these technologies still struggle with noisy, ambiguous, or overly concise responses. Efforts continue to enhance these methods through more sophisticated models, larger training datasets, and respondent-assisted coding processes. However, technical developments often overlook a crucial point: the quality of occupational data—whether for input or training—is fundamentally shaped by how survey questions are phrased.

As noted by the US Census, the wording of the occupational questions impacts the quality of answers and can hinder the quality of manual coding \citep{Martinez2017}. It is plausible that the same influence applies to automatic coding processes. However, the literature on automatic occupational coding has largely neglected the role of question wording and data origin, despite its potential to affect the performance of coding algorithms. Understanding how the framing of occupational questions influences both the richness of the data and its suitability for automatic classification is critical to improving the accuracy of these methods.

This study addresses this gap by examining the effect of occupational question framing on automatic coding. Given that question wording can vary widely and may be influenced by cultural or linguistic factors, we conducted an experiment using two question formats commonly employed in German surveys. Specifically, we address two research questions: Firstly, how does the framing of occupational questions affect the variability and richness of the responses? Secondly, do different question wordings produce information that is equally conducive to automatic coding? Our findings offer practical insights for researchers and practitioners, emphasizing the importance of considering the original question formulations when assessing the performance of coding algorithms.

\section{Background}
The quality of any automatic coding system hinges on the wording of the occupational questions that generate the input data.  In what follows, we briefly trace how survey designers have formulated these questions, summarise empirical evidence on their impact, and situate recent advances in automatic coding within that context.

\subsection{Occupational Question Framing}

Efforts to record occupations in a comparable form began in national censuses of the late 18th – 19th centuries \citep{whitby2020sum}. For instance, the 1850 U.S. Census simply asked for each male person’s “profession, occupation, or trade,” but enumerators relied on a 156-word instruction sheet that grew to more than 2,500 words by 1900 \citep{ipums_usa_v15}.  The International Labour Organization (ILO) raised a concern in 1949: "Since many of the problems [...] are due to the vagueness or insufficiency of the information furnished, particular attention should be paid to the formulation of the questions referring to the occupation" \citep [p. 91]{ilo1949isco}. Forty years later, they reiterated the importance of well-designed and tested occupational questions, recommending the use of two questions — job title and main duties \citep{ilo1987}. Yet formal guidance on how to phrase those questions did not appear until 2012, and even then without methodological backing \citep{ilo2012isco}. Without clear standards, survey designers have adopted a wide array of wording strategies for occupational questions.

From the 1950s until 2024, the U.S. Census and the American Community Survey (ACS) asked between one and three occupational questions, altering the first only once, in 2018, from “What kind of work was he doing?” to “What was this person's main occupation?” \citep{ipums_usa_v15}. Germany, by contrast, tried sixteen different formulations across the Census and the Mikrozensus (see Table 1A in the Appendix) \citep{forschungsdatenzentrum_vz1970, forschungsdatenzentrum_mikrozensus, zensus_1987, zensus2011, zensus2022}. A broader review on the 33 studies in the United States and Europe \citep{Tijdens2014} showed that precise wording of the first question varied although most asked for an “occupation/job title”.  In contrast, German panel studies  prefer "occupational task"-oriented wording \citep{Schneider2022}, and the Swiss Household Panel revised its German-language item in 2014 while leaving the French, Italian, and English versions untouched \citep{Tillmann2021}.  Even a single multinational Labour Force Survey (LFS) is inconsistent: the 2024 Swedish one asked  "Into which occupation would you classify your work?", whereas LFS Luxembourg in 2021 - "Which occupational task did you perform during the reporting week? Please state the exact title," and LFS Portugal in 2021 -"What is your occupation? Please be as comprehensive and detailed as possible and describe the functions or tasks you perform" \citep{EU-LFS-questionnaires}.

In short,  nearly 200 years after occupational questions became part of official statistics their wording remains heterogeneous - across countries, across surveys and within the same survey type. The ILO's calls for standardisation have not produced convergence; current manuals still leave precise phrasing to individual users.  The next section considers what these divergent formulations mean for occupational coding.

\subsection{Question framing and data quality}

Although most respondents understand a question about their “main occupation or title,” the wording is challenging for those with multiple jobs, no formal title, or an industry-specific label that masks what they actually do \citep{ACS2014}.  Early ISCO documentation had already catalogued typical shortcomings, including vague responses such as “worker” or “owner” and inflated titles (e.g., a cashier presenting themselves as an accountant) \citep{ilo1949isco}. If left unclarified, such answers reach coders in unusable form. According to  \citet{geis2000stand}, up to one-quarter of entries cannot be coded to the maximum level of detail, around one-tenth are so generic that coders must guess \citep{Ganzeboom2010}. In addition coders from different agencies agree on only half of the codes, due to both systematic and random errors \citep{Massing2019}.

There have been both theoretical and practical efforts to improve the quality of occupational data. For example, \citet{geis2000stand} building on \citet{pappiSozialstrukturanalysenMitUmfragedaten1979}, former leaders in Germany’s occupational classification efforts, argued that coders first need to know what a person does, not what the job is called.  He therefore proposed a three-step module for coding into ISCO: (i) main task, (ii) clarifying probe, and only then (iii) job title.  This task-first logic was later adopted in the Federal Statistical Office’s standards \citep{demographische2024standards}, though asking three questions can be overly time-consuming \citep{Schneider2022}. Shortened versions are widely used. One option, recommended by \citet{zuell2015}, is to merge main task and clarifying probe into a single question before asking for a job title. An alternative is to ask for a job title in a first question before, optionally, requesting information about the job tasks; see Table \ref{tab:appendix-1B} in the Appendix for example wordings from several large German surveys. Additional refinements such as larger answer boxes \citep{Martinez2017}, illustrative examples \citep{Massing2019}, and targeted interviewer training \citep{cantor1992} have been tested with mixed success.

Current research points to a trade-off when designing occupational questions for successful coding. Because respondents are unsure which aspects of their work matter most, many supply only a job title \citep{KKM20,Christoph2020,Massing2019,Tijdens2014}. However \citet{Martinez2017} found that asking “What is [Name's] occupation?” is expected to provide more specific answers than “What kind of work does [Name] do?”. When it comes to the follow-up question detailing tasks and duties, guiding examples can encourage respondents to provide more details \citep{Massing2019, Martinez2017}, yet several studies show that example-prompted responses can also become longer, and harder to classify \citep{conrad2016classifying,Massing2019,KKM20}. According to \citet{cantor1992}, coders find that longer occupational duties and tasks descriptions are more likely to include contradictory information, making manual coding less reliable. However, \citet{Martinez2017} and \citet{Massing2019} argue that the issue is more nuanced, as some respondents provide better answers that are easier to code when offered examples.

\subsection{Automatic Coding}

Despite the continued reliance on human coders, automatic solutions have become a routine part of occupational data co-production in both general surveys and official statistics production \citep{sebastian_kocar_occupation_2025}. These automatic solutions vary along several dimensions: the type of technology applied, interactive coding during data collection vs. post-survey data processing,fully automated vs. computer-assisted coding systems, and the type of occupational information on which coding is based.

First, early automatic coding relied on dictionary-based approaches \citep{lyberg_automated_1992}. These were followed by more complex rule-based systems that allowed for partial matching, incorporated common abbreviations, spelling errors, or accounted for the absence of certain words in occupational descriptions. CASCOT, developed by \citet{jones2004cascot}, is one of the most well-known multilingual text-matching algorithms of this type. Similar dictionary-based approaches have been used in the German Mikrozensus \citep{bernhard_hochstetter_berufsklassifizierung_nodate}, or commercial tools developed by Verian \citep{cataneda_russel_collecting_2024}. Advances in computational capacity enabled the development of more complex methods, ranging from ensemble classifiers such as SOCcer \citep{Russ2016}, to machine-learning-based methods \citep{schierholz2021machine}, transformer models such as DistilKoBERT \citep{kim_occupation_2024}, and recently large language models such as SOCBot \citep{sturgisSOCbotUsingLarge2025}. These developments have not replaced earlier methods; rather, dictionary- and machine-learning-based approaches remain in active use.

Second, occupational information is frequently collected through surveys, although other sources, such as visa applications or job search platforms, exist. Coding can occur after data collection is completed (for example, when paper census questionnaires are processed post-survey), or during the data collection process itself, as in interactive survey designs \citep{schierholz2018occupation, sebastian_kocar_occupation_2025, sturgisSOCbotUsingLarge2025}. Some tools, such as CASCOT or OccuCoDe \citep{Simson2023}, can be deployed in both settings.

Third, automatic solutions vary by the degree of automation. Fully automatic systems assign the most likely occupational code without human intervention, whereas semi-automatic systems provide suggestions (or a look-up list) that are reviewed and confirmed by an expert coder or by respondents themselves. Fully automatic solutions include PROCODE \citep{savic_procode_2020} or the LLM-based tool by \citet{bach_job_2025}, although many semi-automatic tools (such as CASCOT, OccuCoDe, or SOCcer) can also be used in a fully automatic mode.

Finally, automatic coding tools differ by the type of occupational information they are designed to process. Despite advances in computational capacity, many existing systems are optimised for short textual inputs rather than longer descriptions. Short answers (typically responses to questions such as “What is your job title?”) are commonly used to generate coding suggestions, while follow-up questions about tasks and duties are often reserved for office coding \citep{sebastian_kocar_occupation_2025}. Tools such as CASCOT, OccuCoDe, Verian, PROCODE, SOCBot, Klassifikationsserver by Mikrozensus, primarily rely on short inputs and may not reliably process longer, more detailed descriptions of tasks and duties. A smaller number of tools, including SWEEP \citep{sebastian_kocar_occupation_2025} and SOCcer, are able to incorporate additional information such as tasks or industry alongside job titles.

So far, debates about automatic codability (the likelihood of assigning the correct code) are largely centred on the technical specifics of the algorithms. Yet we still lack empirical evidence that the question formats feeding these algorithms yield occupational data of comparable quality, leaving open whether input design, rather than algorithmic sophistication alone, governs the likelihood of assigning the correct code.

\subsection{Research Gap, Research Questions, Hypotheses}

Against this background, several inconsistencies can be observed in the existing literature on occupational data collection. Although occupational surveys have a long tradition, the literature offers no consensus on the optimal wording of open-ended occupation questions. Prior studies show that small variations in phrasing can alter the specificity of responses and, consequently, the effort required for manual coding. Adding illustrative examples typically prompts respondents to write longer answers, yet these more elaborate texts can further complicate coding. What remains largely unexamined is how these design choices affect the structure of occupational information in ways that are relevant for automatic coding, including their potential implications for the lexical material such tools are able to exploit.

Our research addresses these gaps by investigating the following research questions: What is the variability in responses due to different occupational question framings? Do these different framings yield similarly rich information? Do different question framings produce information that is equally conducive to automatic coding?

To answer these research questions, we designed, conducted, and replicated a survey experiment in Germany using two versions of occupational questions commonly found in German surveys. We then tested the responses using two automatic coding tools available for the German language. A focus is on the starting question because the coding tools we use were not developed to process answers from a sequence of occupational questions. Based on the collected data, we formulated and tested the following hypotheses:

\textbf{H1:} Both automatic tools (CASCOT and OccuCoDe) will show a higher codability rate when coding responses to job titles compared to occupational task.

\textbf{H2:} The question about job titles will encourage job titles, leading to less linguistic diversity in the responses. In contrast, the question about occupational tasks will result in a broader range of linguistic expressions, with respondents providing both occupational titles and descriptive terms, thus increasing linguistic diversity.

\textbf{H3a:} Responses to the question about detailed tasks that include a guiding example will be longer than those to a question without an example.

\textbf{H3b:} Responses to the question about detailed tasks with a guiding example will exhibit greater linguistic diversity than those provided without an example.

\section{Methodology}
This section provides a detailed explanation of our survey experiment and its replication.

\subsection{Data} We conducted an online experimental study managed by Forsa and simultaneously replicated it using the panel managed by Bilendi. Both firms are commercial survey-research companies that maintain respondent panels. Forsa operates a probability-based panel, recruiting respondents via a random-digit dialling probability sample and subsequently inviting them to participate in specific studies \citep{Guellner2004, Schnell2024}\footnote{Forsa confirmed that occupation is not asked as a permanent panel profile variable; while panellists might have been asked about their job in other studies, those data remain study-specific, so no systematic occupation-related priming could influence our experiment.}. Thus, while it is expected to suffer less from self-selection bias, we recognize that other biases, such as differential response rates, may occur. Bilendi, by contrast, is an entirely opt-in panel (non-probabilistic). The Forsa experiment was conducted in Germany between December 7 and 13, 2022. The replication study, conducted by Bilendi, took place over a shorter period, from December 7 to 9, 2022. Both surveys were administered via the Unipark platform, ensuring an identical user experience across both panels. Both companies compensated respondents with digital vouchers for their participation.

\subsection{Experimental Design} Our split-ballot experiment, along with its replication, consisted of two sequential open-ended questions, as illustrated in Figure 1 (the original German version is shown in Figure 1A in the Appendix). Respondents were randomly assigned to see one of the two experimental conditions, ensuring a balanced distribution across versions of two questions.
\begin{figure}[htbp]
\centering
\raisebox{-0.5\height}{\includegraphics[width=\textwidth]{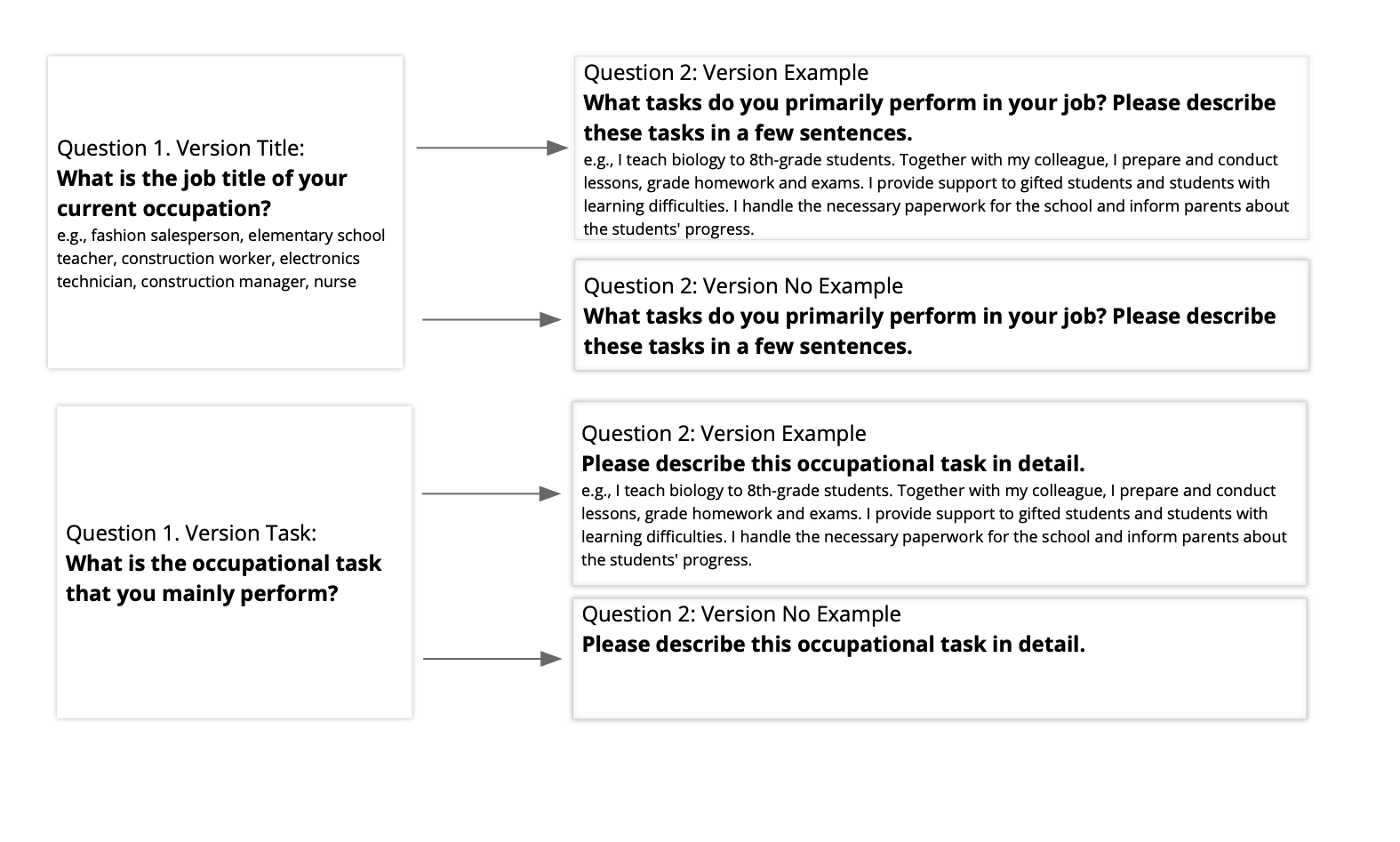}}
\caption{\label{fig:experiment}Overview of the survey experiment and replication study.}
\par 
{\scriptsize
\noindent 
\textnormal{
Experimental design illustrating the split-ballot approach. Respondents were randomly assigned to one of two versions of Question 1 ("Title" or "Task") and, independently, to one of two versions of Question 2 ("Example" or "No Example"). This design tests the effects of each question version separately without examining interactions between the questions.\\[-0.5em]}
}
\end{figure}

The first question involved two different versions of the occupational question commonly used in German surveys (see Table 1B in the Appendix):

\begin{itemize}
\item Question 1"Title": "What is your current job title?" (German: “Welche Berufsbezeichnung hat Ihre gegenwärtige Tätigkeit?”), currently used in the Mikrozensus.
\item Question 1"Task": "What is the occupational task that you mainly perform?" (German: “Welche berufliche Tätigkeit üben Sie derzeit hauptsächlich aus?”), apart from very minor deviations this question text is generally recommended by the German Federal Statistical Office and other institutes \citep{demographische2024standards}. Many large panel surveys adopt the task-focused phrase ``berufliche Tätigkeit'' in their questions, though typically these surveys include job titles as examples or request a detailed description of the occupational tasks. 

According to \citet[pp. 104]{geis2000stand}, the Question 1"Task" followed by question 2 (“Please describe this occupational task in detail.”) is recommended to collect information about the work (tasks and duties) performed by a respondent. We note, however, that the German question text is rather hard to translate to English. Many respondents may just understand the first question as an unusual or formal way of asking “What is your main occupation?”, not noting the intended minor emphasis on which task they do in their job.
\end{itemize}

Respondents were randomly assigned to one of these two versions. Since both questions were framed in the present tense (inquiring about the current job title or primary occupational activity), respondents who were retired or not employed were given the option to select a separate category: "I am retired" or "I am not working," rather than providing a text response. Approximately 40\% of respondents in both surveys selected this option.

The follow-up question asked respondents to provide further details about the tasks and duties they performed in their occupation. Half of the respondents were shown a detailed example of a response: "\textit{I teach biology to 8th-grade students. Together with my colleague, I plan and prepare the lessons, grade homework, and exams. I give extra tuition to gifted pupils and pupils with learning difficulties. I do the necessary paperwork for the school and inform the parents about the students’ progress.}" The other half received no example. Although the exact wording of the second question varied slightly depending on the phrasing of the first question, in this article we focus solely on the presence or absence of the example and treat these slight differences as insignificant.

\subsection{Analysis}
This section explains the measures and analytical tools we use to test the hypotheses. All tests are applied separately to the experimental and replication study. 

\textbf{H1. Question 1: Job Titles/Occupational Tasks Automatic Codability} We use two readily available automatic coding tools for the German language, CASCOT and OccuCoDe, that can code answers into national German (KldB 2010) and international (ISCO-08) classifications. We code answers after the survey has been conducted, however OccuCoDe has a functionality to be embedded directly in the survey for coding during the survey. When either algorithm assigns a code to a job title or an occupational task, it produces a confidence score to indicate the level of certainty that a given occupation should receive a specific code. For example, OccuCoDe will assign an answer “surgeon” (Chirurg) to the German classification KldB 2010 code 81434 (Surgeon) with a confidence score of 0.8. The second most likely choice will be the code 81474 (Dentist) with a confidence score of 0.05 etc. CASCOT will assign the same job to the 81434 (Surgeon) with a confidence score of 70. The second most probable job code will be 81332 (Surgical technical assistant) with a confidence score 20, and so on. Both tools aim to offer high-quality suggestions and warn against relying on suggestions that have a low score. In practice, it means that if a person answers “Civil servant” (Beamter), OccuCoDe will suggest that the most likely job category is “Administrative assistant” (Verwaltungshelfer/in) with a confidence score of 0.46, CASCOT will offer “civil servant working at a stud farm” (Beamter/Beamtin Landgestüte) with the score 52. Both suggestions lie below the tools’ recommended minimum confidence score levels (0.535 for OccuCoDe (possible range: 0-1) \citep{Simson2023} and 64 (possible range: 0-100) for CASCOT \citep{jones2004cascot}). As a result, the job “civil servant” is considered not precise enough for automatic coding and will require a manual coder's effort. Because each case that falls under the confidence cut-off is automatically flagged in the output file, the automatic tools are reliably error-prone: generic responses such as "civil servant" surface in a predictable way, allowing for targeted follow-up, a level of traceability rarely attainable with human coders. We will call occupations that can be confidently coded using automatic tools “easy-to-code” (CASCOT: confidence scores 64-100; OccuCoDe confidence scores 0.535-1); the remaining ones will be called “hard-to-code”(CASCOT: confidence scores 0-63; OccuCoDe confidence scores 0-0.5349) and include some non-codable occupations (score 0). Importantly, human coders will more often agree with easy-to-code answers and less often agree with hard-to code ones, meaning that the accuracy from hard-to-code answers is believed to be lower. We do not aim to compare the coding decisions of the two tools. Instead, we are only interested in whether more answers to the job title question (Question 1"Title") fall under the “easy-to-code” category than the ones asking about occupational task (Question 1"Task").

In order to evaluate if the origin of data affects automatic codability, we select the code with the highest confidence score for each answer and calculate two binary variables. Outcome 1 indicates that a given answer's highest confidence score is equal or above 0.535 (OccuCoDe) -- easy-to-code; otherwise, 0 -- hard-to-code. The second variable is coded as 1 (easy-to-code) when CASCOT gave a score above 64 to a response; otherwise 0. We apply $\chi^2$ test (p= 0.05) to test the Null-hypothesis that there is no difference in the rates of easy-to-code answers between "job title" and "occupational task" versions.

\textbf{H2: Question 1: Job Title/Occupational Task linguistic diversity} To evaluate lexical diversity in the responses to the Question 1: job title/occupational task question we first defined six, partly overlapping, lexical categories: nouns, nouns that represent an occupational title (ones that end with -mann, -frau, -er, -erin, -in, -eur, -e, -ist, -wirt), verbs, adjectives, stop words (German stop words like und, ich, der, die, das, etc.), derived nouns (nouns that are derived from verbs and have endings like -ung, -heit, -keit, -nis, -schaft). We then prompted the Azure OpenAI GPT-4 model solely for part-of-speech tagging, assigning each response a binary indicator that recorded whether at least one token belonging to each category was present (1) or absent (0).  For example, "Schulleitung" (school leadership/direction) was coded as noun - 1, verb - 0, adjectives - 0, stop words - 0, derived noun - 1. We evaluate the difference in answers using Exact Fisher’s test (= 0.05) and tested the sets of Null Hypotheses that there is no difference in usage of different lexical groups between "job title" and "occupational task" versions. 

Second, we preprocess the responses by tokenizing, removing stop words, and applying stemming. We then measure linguistic richness using the Type to Token Ratio (TTR), which evaluates the number of unique words relative to the total number of words. The TTR ranges from 0 to 1, with higher values indicating a greater proportion of unique words \citep{deBoer2014, Gavs2022}. We calculate separate TTR for answers to job titles and professional activities, measure the difference TTRdif = TTRprofessional activity - TTRjob titles, and apply bootstrapping (N=1000) to estimate whether the difference is statistically significant (= 0.05). 

In addition, following \citet{Martinez2017, Struminskaya2015, meitinger2021}, we compare job title responses with professional activity using the length of answers (in words and characters). We apply a two-sided Mann–Whitney U (MWU) test test to measure the difference in length, as well as Levene's test to measure the variability in answer length. 

\textbf{H3: Question 2: Detailed tasks that include/exclude a guiding example} 
Since both CASCOT and OccuCoDe aimed to handle only a single free-text field, we provided \emph{only the answers to Question 1} to these automatic-coding tools. Answers to Question 2 are analysed exclusively for surface characteristics (length, lexical diversity) and never used to measure codability.

First, we compare the length of answers (in characters and words) to the Question 2 (detailed tasks) with the example and without the example \citep{Martinez2017, Massing2019}. We apply the one-sided MWU test and Levene's test to measure the difference in answer length variation. Second, we used the same lexical categories defined in Hypothesis 2: nouns, nouns that represent an occupational title (ones that end with -mann, -frau, -er, -erin, -in, -eur, -e, -ist, -wirt), verbs, adjectives, stop words (German stop words like und, ich, der, die, das, etc.), derived nouns (nouns that are derived from verbs and have endings like -ung, -heit, -keit, -nis, -schaft). We again used Azure OpenAI to classify each response into these six categories.

However, because responses to Question 2 (tasks and duties) are usually longer than a single word, we counted the number of words in each category. For example, the sentence "Ich begleite einen Schüler mit Asperger-Syndrom an Gymnasium" (I accompany a student with Asperger syndrome at Gymnasium) was tagged as: nouns – 3, verbs – 1, adjectives – 0, stop words – 4, derived nouns – 0. use a $\chi^2$ test (= 0.05) for the $H_0$ that there is no difference in the count of words between versions with example and without example. Third, as we did for Hypothesis 2, we also measured lexical diversity using TTR difference between respondents who answered the question with an example and without one.

\section{Results}
\subsection{Descriptive Statistics}
We received 840 responses from the two agencies after excluding everyone who explicitly selected the option “I am retired, or I am not working.” As recorded in Table 1, two respondents did not answer a question (Hard non-response), and 63 respondents (7\%) provided an answer that was classified as a Soft non-response (gibberish, irrelevant answer etc.) and also excluded from further analysis.  All remaining responses were spellchecked and (where appropriate) brought to the same form (for example, Kfm Angestellster, kfm. Angestellter, kaufmännischer Angestellter --> kaufmännischer Angestellter) in order to exclude the effect of spelling on codability \citep{bao2020occupation}. The final number of observations with valid responses is 775. Respondents, on average, used 1 (Bilendi) to 1 (Forsa) words or from 15 to 16.7 characters to describe their job title (Quesions 1"Title") and from 1 (Bilendi) to 1 (Forsa) words or from 15 to 17 characters to answer the question about the occupational task (Question 1"Task").

{\scriptsize
\begin{longtblr}[
  caption={Descriptive Statistics of Response Patterns for Job Title (Title) and Occupational Task (Task) experimental conditions.}, 
  label={tab:descriptive}, 
]{
  colspec={p{6cm}|c|c|c|c}, 
  rowhead=1,
}
\SetCell[c=1]{c}  & \SetCell[c=2]{c}\textbf{Forsa} & & \SetCell[c=2]{c}\textbf{Bilendi} & \\\hline

{Category} & \parbox[c]{2cm}{\centering{Task}} & \parbox[c]{1.5cm}{\centering{Title}} & \parbox[c]{1.5cm}{\centering{Task}} & \parbox[c]{1.5cm}{\centering{Title}}\\
Hard nonresponse (HNR) & 1 & 0 & 1 & 0 \\ 
Soft nonresponse (SNR) & 12 & 14 & 23 & 14 \\ 
Valid response & 189 & 147 & 224 & 215 \\ 
Mean length of the answer (N characters) & 17 & 17 & 15 & 15 \\ 
Median length of the answer (N characters) & 14 & 16 & 13 & 14 \\ 
Mean length of the answer (N words) & 1 & 1 & 1 & 1 \\ 
Median length of the answer (N words) & 1 & 1 & 1 & 1 \\ 
\end{longtblr}
}

Descriptive statistics in Table 2 provide a summary of answers to the detailed tasks question where an example was provided compared to the question where no example was provided. Answers to the question that included an example are longer: the average answer is between 66 and 104 characters long or between 8 (Bilendi) and 13 (Forsa) words long. Answers to the questions where the example was not provided are between 47 and 57 characters long or between 6 and 7 words long. Respondents to the Bilendi-administered survey provided shorter answers than did respondents to the Forsa-administered one. 

{\scriptsize
\begin{longtblr}[
  caption={Descriptive Statistics of Response Patterns for questions with the example and questions with no example.},
  label={tab:descriptiveexample},
]{
  colspec={p{6cm}|c|c|c|c}, 
  rowhead=1,
}
\renewcommand{\arraystretch}{0.02} 
\SetCell[c=1]{c}  & \SetCell[c=2]{c}\textbf{Forsa} & & \SetCell[c=2]{c}\textbf{Bilendi} & \\ \hline
{Category} & \parbox[c]{2cm}{\centering{Example}} & \parbox[c]{1.5cm}{\centering{No Example}} &  \parbox[c]{1.5cm}{\centering{Example}} & \parbox[c]{1.5cm}{\centering{No example}}\\

Hard nonresponse (HNR) & 9 & 5 & 6 & 5 \\ 
Soft nonresponse (SNR) & 16 & 11 & 27 & 34 \\ 
Valid response & 164 & 159 & 192 & 212 \\ 
Mean length of the answer (N characters) & 104 & 57 & 66 & 47 \\ 
Median length of the answer (N characters) & 78 & 43 & 48 & 33 \\ 
Mean length of the answer (N words) & 13 & 7 & 9 & 6 \\ 
Median length of the answer (N words) & 9 & 5 & 6 & 3 \\ 
\end{longtblr}
}

\subsection{Codability of answers}
 We tested the codability of answers to the questions about job titles and occupational task using the two automatic occupational coding tools introduced in the Methods section: OccuCoDe and CASCOT. The results show that over 50\% of responses are hard-to-code for the automatic tools regardless the question respondents answered. The range of codability difference was similar for both Forsa and Bilendi samples. For example, as seen in Table 3,  40\% of responses received by Forsa study to the question about occupational task and 48\% of answers to the question about job titles were easy-to-code when OccuCoDe was used. The rest are hard-to-code answers and need to be classified manually. It is important to note that answers flagged by automatic tools as “hard-to-code” pose a similar challenge for manual coders and typically require additional context (e.g. information about industry, detailed duties, education, or type of employment) to assign a detailed code unambiguously.
 
{\scriptsize
\begin{longtblr}[
  caption={Summary of Automatically Codable Responses and Statistical Tests},
  label={tab:summary},
]{
  colspec={p{2.3cm}|p{1.1cm}|p{1.1cm}|p{3cm}|p{0.6cm}|p{3.4cm}},
  rowhead=2
}
\SetCell[r=2]{c}\parbox[c]{2.3cm}{\textbf{Tool/Study}} & 
\SetCell[c=2]{c}\parbox[c]{2cm}{\textbf{Automatic Coding Rate}} & &
\SetCell[r=2]{c}\parbox[c]{3cm}{\textbf{$\chi^2$-Test for Codable Answers ($\alpha$ = 0.05)}} &
\SetCell[r=2]{c}\parbox[c]{0.6cm}{\textbf{N}} &
\SetCell[r=2]{c}\parbox[c]{3.4cm}{\textbf{Difference in Codability (Title - Task)}} \\
& Task & Title & & & \\ \hline
OccuCoDe-Forsa   & 40\% & 48\% & $\chi^2(1) = 1.81,\ p = .18$ & 336 & 0.079; 95\% CI: [-0.03 , 0.19] \\ 
OccuCoDe-Bilendi & 39\% & 46\% & $\chi^2(1) = 2.04,\ p = .15$ & 439 & 0.072; 95\% CI: [-0.02 , 0.17] \\ 
CASCOT-Forsa     & 30\% & 31\% & $\chi^2(1) = 0.04,\ p = .83$ & 336 & 0.017; 95\% CI: [-0.09 , 0.12] \\ 
CASCOT-Bilendi   & 33\% & 34\% & $\chi^2(1) = 0.01,\ p = .92$ & 439 & 0.009; 95\% CI: [-0.08 , 0.10] \\ 
\end{longtblr}
}

Hard-to-code responses included terms such as clerk, employee, self-employed, worker, consultant, and salesperson, among others. These are generic and ambiguous answers for which multiple occupational categories could be equally appropriate, as indicated by their low confidence scores. Both CASCOT and OccuCoDe showed a higher proportion of easy-to-code responses when participants answered the job title question. However, according to a series of chi-squared ($\chi^2$) tests, there was no statistically significant difference in the proportion of easy-to-code responses between Question 1"Title" and Question 1"Task". The 95\% confidence intervals nonetheless indicate the range of effects still compatible with our data: for OccuCoDe (Forsa) the true difference could be anywhere from –3 to +19 percentage points, and for CASCOT the bands are –9 to +12 points in the Forsa sample and –8 to +10 points in Bilendi. Translating those limits into workload, switching to the job-title wording could, in the worst case (-9 pp), raise the manual-coding load by roughly 9\% (about 12600 extra cases in a labour-force survey with 140000 interviews, the size of the German LFS). In the best-case (+19 pp) it could let about 26600 more answers be coded automatically. Since these bounds are wide, a larger study is needed before survey practitioners can choose the wording with firmer statistical confidence.

\subsection{Linguistic diversity of job title/occupational task answers}
The evidence from Figure 2 indicates that most of the answers followed the same pattern: irrespective of the formulation of the question asked, almost all respondents answered with a job title. Fisher's Exact Test confirms that there is no statistical difference between the types of words used to answer the first occupational question, whether asking about job title or occupational task. The result is identical to the main experiment and its replication. (\textit{Forsa}\textsubscript{p-value} = 0.89, \textit{Bilendi}\textsubscript{p-value} = 0.56).

\begin{figure}[htbp]
\centering
\raisebox{-0.5\height}{\includegraphics[width=0.8\textwidth]{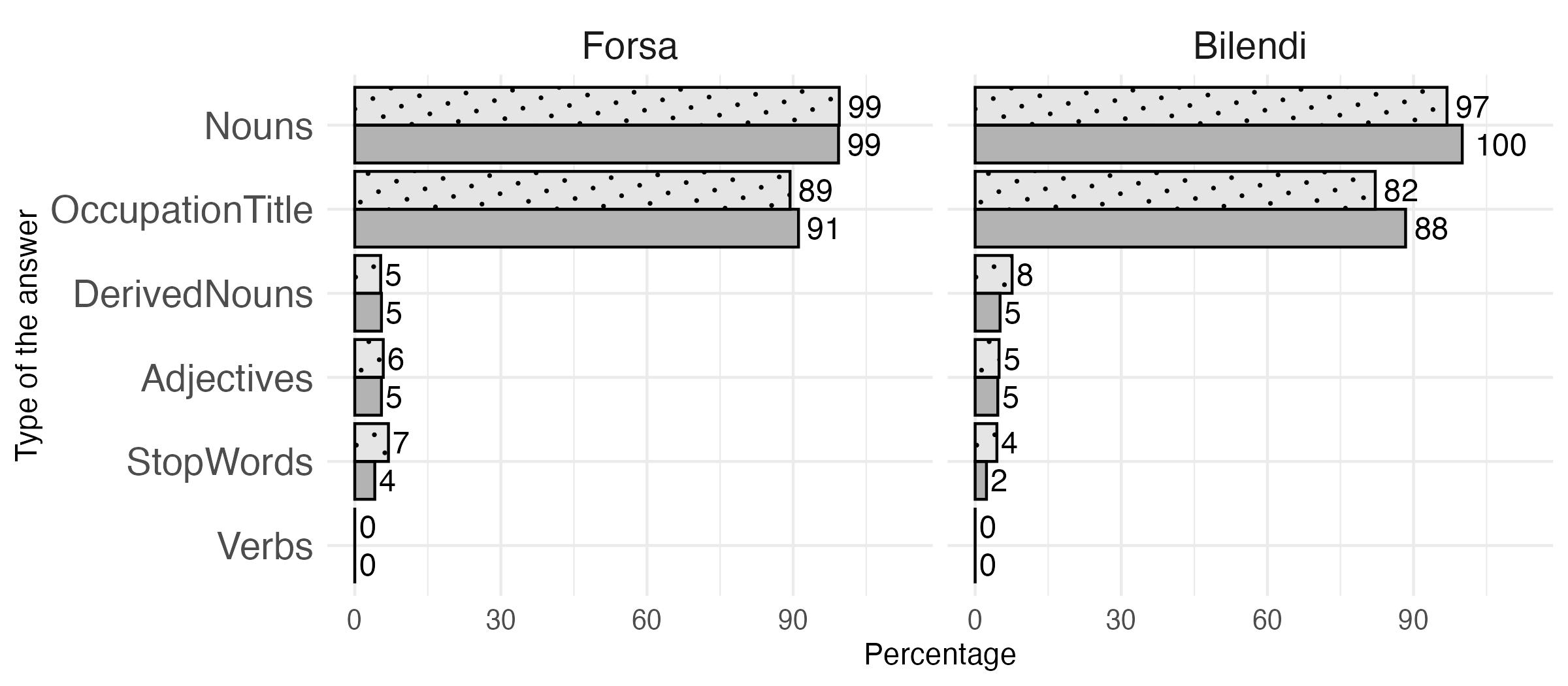}}
\caption{\label{fig:wordtypes.1} Distribution of Word Types in Responses to the Question 1.}
{\scriptsize
\noindent 
\textnormal{
The figure shows the percentage of different types of words (e.g., nouns, adjectives) that respondents used in their answers to the Question 1 variations. Solid grey bars show the percentages for answers to Job Title question (Question 1"Title"), while dotted grey bars show the percentages for answers to Occupational Task question (Question 1"Task").\\[-0.2em]}
}
\end{figure}

Next, we formally tested the linguistic diversity using TTR (type-to-token ratio).
The results show that respondents used fewer unique words when answering the question about occupational task than job title TTR$_{\text{diff}} = -0.1$ for Forsa and TTR$_{\text{diff}} = -0.03$ for Bilendi. However there is no statistical evidence (bootstrapped results ($N = 1000$)) that the variety of words used to answer job title question is different from that of occupational task answers Forsa: $t = -0.1$, $\text{bias} = 0.05$, $\text{std.error} = 0.05$. Bilendi: $t = -0.03$, $\text{bias} = 0.01$, $\text{standard error} = 0.04$]. The results of the MWU test as well as Levene's test are consistent for experiment and replication and indicate that there is no statistical difference between the length of responses [Bilendi: $F \text{(1,437)} = 0.03$, $\text{Pr}[>F] = 0.9$, $\alpha = 0.05$; Forsa: $F \text{(1,334)} = 1.16$, $\text{Pr}[>F] = 0.29$, $\alpha = 0.05$.] (See Figure 2A in the Appendix). 

The lexical analyses above show that respondents frequently use job-title nouns, regardless of whether we ask for a title or a task. Yet Table 3 makes it clear that, despite this surface similarity, the two question wordings do not yield equally codable answers. To unpack this discrepancy, we condensed each condition to a set of unique strings, retaining only one instance of every distinct job-title response (e.g., the two occurrences of “Employee - Angestellter” in Bilendi–Task count as a single token). The box-plots in Figure 3 provide the resulting codability distributions for CASCOT (left) and OccuCoDe (right). A pattern stands out:

Longer left-tail for the task wording: Occupational task responses generate a higher share of low confidence scores when using OccuCodeDe, which is especially evident for the Forsa sample. By looking at these low-scoring  occupational task answers, we observe a number of generic titles such as “IVDR-Produktspezialist”, "Standortleiter", or “Direktor der Abteilung”, "Inklusionsassistentin", which apparently are less frequent under the job-title condition. Another set of generic responses such as “Beamtin”, “Arbeiter”, “selbständig”, “Kaufmann”, "Ingenieur", “IT-Manager” appear in both formulations and score poorly in each.

In other words, in both samples, the type of linguistic material is similar, but its semantics differs. The task wording elicits more vague descriptions that are hard to automatically code, pulling down the overall confidence scores. This nuance explains why codability diverges even when grammatical composition (job title type of nouns vs. other word classes), length of answers and lexical diversity (TTR) do not.

\begin{figure}[htbp]
\centering
\raisebox{-0.5\height}{\includegraphics[width=0.8\textwidth]{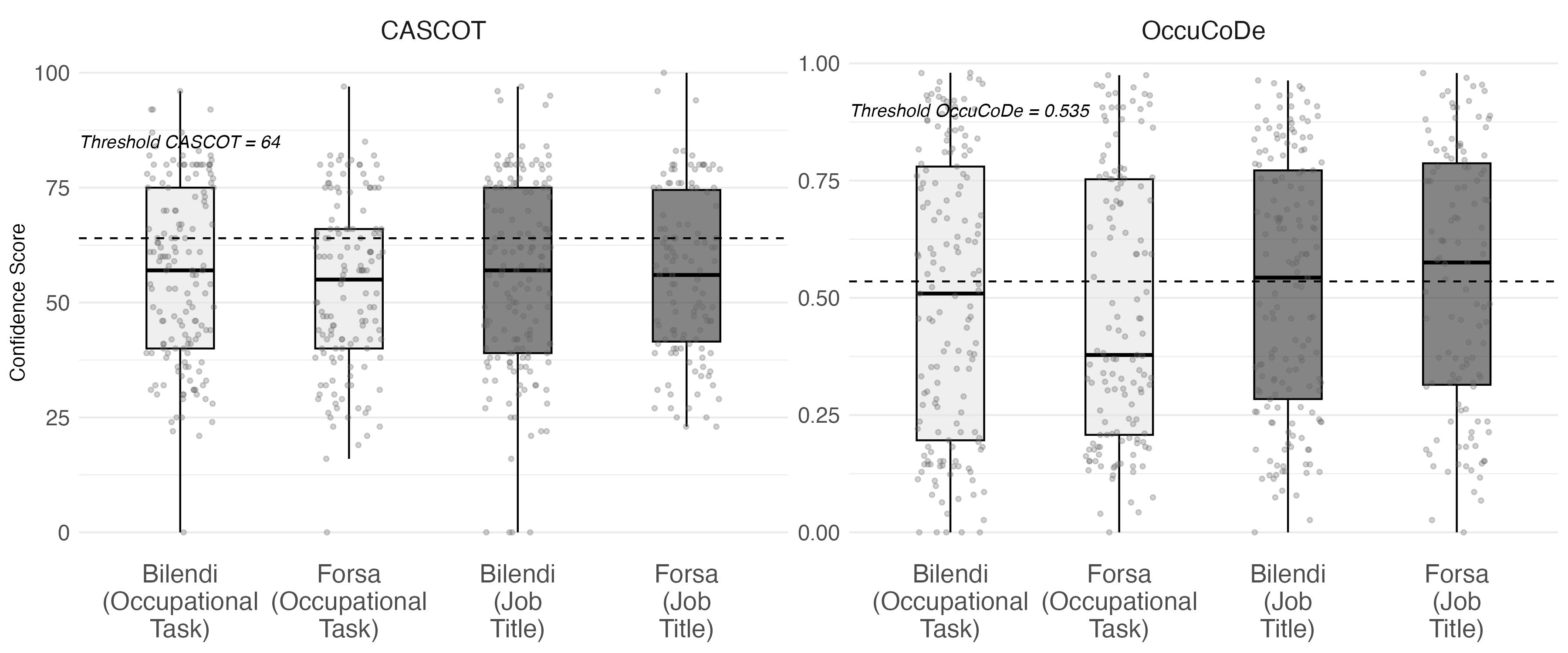}}
\caption{\label{fig:codability} Distribution and characteristics of answers based on length (in characters).}
{\scriptsize
\noindent 
\textnormal{
The figure shows the distribution of codability scores for two Question 1 versions: "Job Title" (dark grey) and "Occupational Task" (light grey). Dashed lines represent a threshold for each of the coding systems ("easy to code" vs "hard to code").\\[-0.5em]}
}
\end{figure}

\subsection{Response patterns to the detailed-tasks Question 2 (example vs. no example)}

The responses to the question with examples are significantly longer than those without an example. Responses were longer in both samples [Forsa: $F \text{(1,321)} = 16.8$, $\text{Pr}[>F]=0.00$], [Bilendi: $F \text{ (1,402)} = 3.8$, $\text{Pr}[>F]=0.05$] (See Figure 3A in the Appendix)

\begin{figure}[htbp]
\centering
\includegraphics[width=0.8\textwidth]{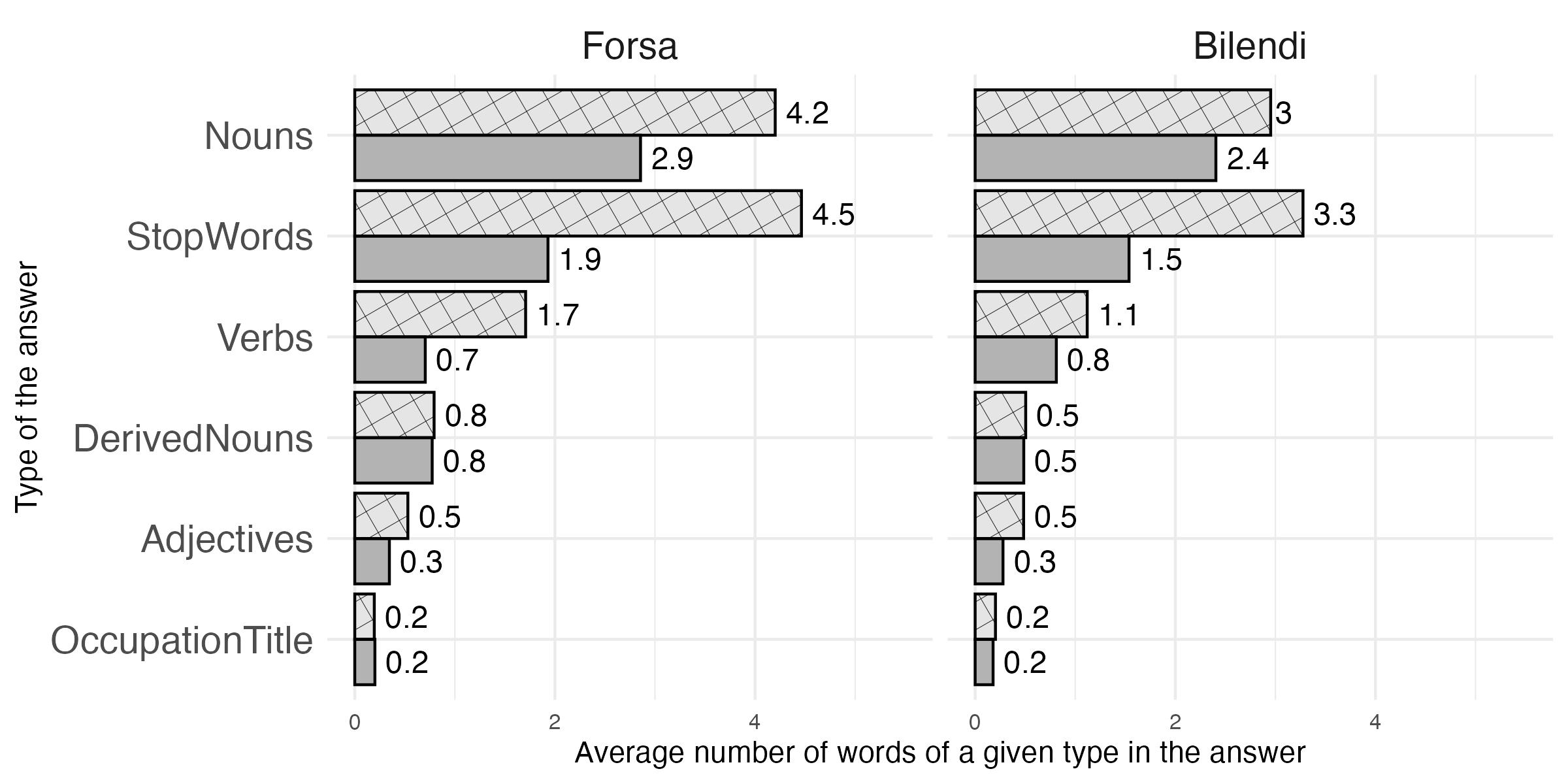}
\caption{\label{fig:diversity.2} Word types in responses to Question 2 (with vs. without example).}
{\scriptsize
\noindent 
\textnormal{
The figure shows the average number of different types of words (e.g., nouns, adjectives) that respondents used in their answers to the Question 2 (Detailed Job descriptions) variations. Patterned bars show the average for answers to the Q2-Example, while solid grey bars show the averages for answers to the question without example (Q2-Example)\\[-0.5em]}
}
\end{figure}

As detailed job descriptions are longer, the pattern of word types also changes (Figure 4). Nouns are still the most frequently used type of word; however, they are no longer occupational titles. Moreover, on average, responses with an example contained on average 2 verbs (Forsa) and 1 verb (Bilendi), whereas responses without an example contained 0.7 and 0.8 verbs, respectively. A $\chi^2$ test confirms that the frequency of words used to answer the question with an example is different from that corresponding to the question with no example (Forsa: $\chi^2$(5) = 58.51, $p = .00$, $\alpha = 0.05$; Bilendi: $\chi^2$(5) = 47.3, $p = .00$, $\alpha = 0.05$).

The specific words used also changed. For instance, the verb “führen/leiten” (“to lead/manage”) appeared only in the example condition. Manual inspection revealed twelve cases in which respondents listed a non-managerial job title (e.g., kindergarten teacher, bank clerk, geriatric nurse, web developer) yet described managerial duties using “führe-/leit-”. Eight of these respondents had seen the example and four had not. For example, one geriatric nurse added:  “I manage a residential area in a retirement home. I coordinate the assignments of my staff, write duty rosters, and liaise with relatives.” (“Ich leite einen Wohnbereich in einem Seniorenheim. Ich koordiniere die Einsätze meines Personals, schreibe Dienstpläne und stehe in Kontakt mit Angehörigen.”)

Finally we assessed lexical diversity using the difference in type–token ratios (TTR). The bootstrapped results show that respondents used essentially the same proportion of unique words whether or not they were shown the example : [Forsa: $t = -0.06$, $\text{bias} = 0.02$, $\text{std.error} = 0.05$; Bilendi: $t = -0.01$, $\text{bias} = -0.004$, $\text{std.error} = 0.05$] 

The results support Hypothesis 3a - responses to the questions with examples were significantly longer. Hypothesis 3b was not supported: despite the increased length and a higher frequency of nouns and verbs that help (for example) identify managerial roles, overall vocabulary diversity remained comparable across conditions.

\section{Conclusion}
For more than a century, questionnaire designers have been thinking about the wording of occupational questions. It is now widely recommended to ask a sequence of two (or more) questions so that as many respondents as possible, even the hard-to-code ones, provide all the information needed for human coding. From this perspective, the goal is to maximize the information content of the complete sequence of questions. Questions at the start of the sequence developed under this perspective frequently steer respondents away from replying with their job titles. Another perspective emerges when considering automatic coding tools that have penetrated occupation coding in practice. Most existing tools rely on short inputs, particularly job titles, and may not reliably process longer, more detailed descriptions of tasks and duties. Since most respondents mention a job title in response to a starting question about their occupation, the information content from this first question is highly relevant for automatic coding approaches. We observe some possible tension between these two perspectives and, in practice, the need to decide whether the starting question should aim to collect job titles.

Our research focused on investigating the impact of occupational question wording on variability and the automatic codability of answers. In order to test that, we conducted a split-ballot survey experiment and simultaneously replicated it. All our findings hold for both studies and show that question framing should be considered when designing, testing or applying automatic coding tools.

First, between 30\% and 48\% of occupational answers, depending on the question formulation, were confidently coded by the two automatic coding tools we tested. Second, both automatic tools performed better when coding responses to the "job title" question compared to the "occupational task" question. This was especially visible for OccuCoDe, where responses to the job title question had 8\% higher codability rate. Third, contrary to our expectations, we found that irrespective of the question formulation, over 80\% of respondents named their job title. However, respondents to the "occupational task" question more frequently provided generic or ambiguous titles that were not reliably codable by the two automatic tools. Fourth, in line with the findings of \citet{Martinez2017}, the answers to the second question about detailed tasks and duties proved to be significantly longer if an example was provided. The response types also varied by question formulation, with answers to the example-based question containing more nouns and verbs (e.g., "I manage," "I lead"). However, examples steered respondents toward using certain words. The average vocabulary diversity per word did not differ between questions with and without examples.

Our study has several limitations. The sample size was too small to detect statistically significant differences regarding the codability rates we observed. While recognizing this limitation, we consider our results worth reporting for several reasons. First, our main and replication studies consistently yielded similar outcomes. Second, although the observed effects were statistically insignificant, their practical relevance may still be noteworthy for survey operations. For instance, the upper ends of the confidence intervals are large enough to matter in day-to-day survey work: if a job title wording is chosen, a 9 pp loss translates into roughly 12600 extra manual codings in a 140000-interview labour-force survey, while a 19 pp gain would let about 26600 more cases be coded automatically.  Nonetheless, due to statistical insignificance, we cannot rule out that these observed effects may differ in magnitude or direction in a larger or different sample. Thus, our results primarily provide a preliminary practical benchmark for future studies and survey practitioners considering automatic coding, rather than definitive evidence.

Moreover, we only tested two question wording bundles currently recommended or used in Germany --- “berufliche Tätigkeit” ("occupational task") without examples, and “Berufsbezeichnung” ("job title") with examples. A larger, fully crossed design  (including a hybrid "berufliche Tätigkeit" with examples and a more sparse version of "Berufsbezeichnung") could disentangle the distinct roles of stem wording and example provision. Future studies might also examine whether task-oriented examples (instead of current job-title examples) enhance coding performance for occupational task questions. Future research should test the item-succession effects. For instance, asking about occupational tasks first might prime respondents to provide more specific job titles in a follow-up question. This hypothesis, implicit in the \citet{geis2000stand} recommendation, remains untested and warrants future research. Additionally, since our samples came from commercial panels, we cannot exclude the possibility that the very experienced respondents chose an easy path, providing minimal answers to open-ended occupational questions. These panels (treated equally in our analyses despite Forsa’s probability-based recruitment and Bilendi’s opt-in nature) may also underrepresent some hard-to-reach or specialized occupations. Future research should replicate this experiment in a non-panel study setting to address the above concerns.

Our findings have several practical implications. The wording of the occupational question appears to impact data quality for manual and automatic coding, even if our evidence is preliminary. As noted by \citet{ACS2014}, asking to name the job title might not be the best option for those with multiple employments or industry-specific jobs, but, according to our findings, the job title is the answer that will be most likely given. Researchers should pay attention to the wording of the occupational question when given the data for training and testing automatic occupational tools. Working with "job title -Berufsbezeichnung" instead of "occupational task -berufliche Tätigkeit" in German context may affect the accuracy of the automatic coding tools without adjustments to the algorithm. The findings might be different for other cultural and linguistic contexts; however, together with \citet{Martinez2017}, our study points out that the origin of the data matters. Future studies should examine causal mechanism that trigger difference in responses.

We note that our findings are most directly relevant to post-survey coding of open-ended responses. Following the recommendations from \citet{geis2000stand} and \citet{demographische2024standards}, many German surveys ask for "occupational task" as a first question, forming a particular German tradition. Their occupational task question bundle is meant for coding into ISCO, but asking three open-ended questions in a survey is time-consuming and a challenge for any automatic coding solution. If there is only time to ask a single question or if the use of automatic coding is desired, our results suggest that the job title question will perform better than a single job task question, due to higher information density in job titles. While we applied OccuCoDe for post-survey coding, the tool (like some other systems) can also be used interactively during data collection, allowing respondents to select from pre-classified categories. Such interactive approaches may change the coding dynamic (for example, respondents can revise their initial answer if the suggested categories are unsatisfactory), and represent an alternative option for researchers designing new surveys.

Next, our study showed that working with only short job title types of answers is limiting. Following \citet{Christoph2020}, open-ended occupational questions do not always make sufficiently clear which aspects of occupational information are required, resulting in many answers that are hard to code. Our results showed that detailed duties questions gave crucial information about some respondents' managerial duties that could not be foreseen from their initial answers. Therefore, researchers in the field of automatic coding could aim not to reach the maximum level of detail with coding job titles but instead to expand allowed input to include detailed task descriptions or industry. 

In addition, our study showed that asking respondents for detailed descriptions of their duties elicited longer responses. Previous research indicates that more extensive input benefited some automatic coding tools \citep{Russ2016, li2023llm4jobs}.  However, we also found that respondents follow the answer pattern and provide longer but similar answers using the same pool of words. Thus, researchers should not expect that their trained machine learning models are easily transferable and with equal performance from one survey to a different one because the respective question formulations in each survey will always influence what exactly respondents answer.

Overall, our study experimentally explores possible variations of German occupation questions that \citet{geis2000stand} and \citet{demographische2024standards} recommend. Although we are not aware of any study that has fully implemented the lengthy three-question sequence proposed in these recommendations, they have been influential in the design of German occupation questions. Consequently, the wordings of the first occupation-related question in different surveys often resemble one another. However, we would like to highlight a number of differences and considerations: Firstly, different German surveys use the terms `Berufsbezeichnung' (job title), `Beruf' (occupation) and `berufliche Tätigkeit' interchangeably. Secondly, many German surveys (e.g., NEPS, GLES, SOEP, FReDA) use the task-focused phrase `berufliche Tätigkeit' from the recommendations. Their questions mention illustrative job titles in the instructions, clarifying how they expect respondents to answer. This is, however, contrary to the original intent of the phrase which was to steer respondents' attention away from job titles and towards the tasks and duties performed in the job. Despite there being no such clarifying examples of job titles in our study, over 80\% of our respondents mentioned a job title when asked about their `berufliche Tätigkeit'. This suggests that many respondents immediately want to answer with a job title when being asked about their occupation, regardless of the exact wording of the question. Thirdly, one line of reasoning expects higher quality from the job title question if a preceding item asks about detailed occupational tasks, while others reverse the question order, asking for job title first. Finally, if there is only time to ask a single question or if the use of automatic coding is desired, our results suggest that the job title question with examples will perform better than a single job task question without examples, due to a higher information density of job titles. While we applied OccuCoDe for post-survey coding, the tool (like some other systems) can also be used interactively during data collection, allowing respondents to select from pre-classified categories. These interactive approaches may alter the coding dynamic (for instance, respondents can revise their initial answer if the suggested categories are unsatisfactory), and offer an alternative option for researchers designing new surveys. We view these differences as a call to action for future research to explore which question formulation yields the best results.

More generally, the results underscore what survey researchers have long known: question wording matters. Building greater consensus about how exactly researchers should ask for and code occupations in their respective surveys will need additional efforts. Especially as machine learning and AI methods spread, and coding tools are transferred between different surveys, it would be well worth it.

\section{Acknowledgments}

This paper is written with the support of the DFG grant 290773872. The work was done [in part] while one of the authors (Frauke Kreuter) was visiting the Simons Institute for the Theory of Computing. We thank Frederic Gerdon for his help in organizing data collection, Claudia Kuhnke for sharing old Mikrozensus questionnaires,  Wiebke Weber and Regina List for their valuable comments and edits. 
We used Azure Open AI GPT-4 for R code suggestions and to format Overleaf tables; to translate and tag answers (as detailed in the Analysis Section). Elicit and ResearchRabbit were used to collect relevant papers for the literature review, Grammarly and EditGPT (mode “proofread” and “thesis”) for spellcheck and proofreading. No text was automatically generated using AI-based tools, and authors take full responsibility for it.


\bibliographystyle{rss} 
\bibliography{example}   

\newpage
\appendix

\section*{Appendix}

\renewcommand{\thetable}{1A}

{\scriptsize
\begin{longtblr}[
  caption={Comparison of Census and Survey Questions in the USA and Germany (1950-2024)},  
  label={tab:appendix-1A},  
]{
  colspec={p{1.5cm}|p{6cm}|p{6cm}},
  rowhead=1
}
\textbf{Years} & \textbf{USA, ACS - American Community Survey
P - population census  \citep{ipums_usa_v15}} & \textbf{Germany, M - Microcensus
P - population Census
\citep{forschungsdatenzentrum_mikrozensus, forschungsdatenzentrum_vz1970, zensus_1987, zensus2011, zensus2022}} \\ \hline
1950-1959 & 
P: What kind of work was he doing? & 
M: What profession does the employed person practice? \newline
M: In which profession is the household member employed? \newline
M: Which job is practiced in this professional activity? \\ \hline

1960-1969 & 
\textit{No change in question wording} & 
M: What activity (profession) is being practiced? \newline
M: What profession do you currently practice, or what profession did you practice most recently? \\ \hline

1970-1979 & 
P: What kind of work was he doing? \newline
P: What were his most important activities or duties? \newline
P: What was his job title? \newline
P: Activity performed. Keyword-like description & 
M: Current activity (profession practiced) \\ \hline

1980-1989 & 
P: What kind of work was this person doing? \newline
P: What were this person's most important activities or duties? \newline
P: What is your occupation or professional activity? & 
M: Current professional activity (occupation practiced) (before 1982) \newline
M: What is your current occupation? (after 1982) \\ \hline

1990-1999 & 
\textit{No change in question wording} & 
M: \textit{No change in question wording} (until 1995) \newline
M: What is your occupation? (after 1995) \\ \hline

2000-2009 & 
ACS: \textit{No change in question wording} & 
M: \textit{No change in question wording} \\ \hline

2010-2019 & 
ACS: \textit{No change in question wording} (before 2018) \newline
ACS: What was this person's main occupation? \newline 
ACS: Describe this person's most important activities or duties. (after 2018) & 
M: \textit{No change in question wording} (until 2010) \newline
M: Enter the job title for your professional activity (2011) \newline
M: State your job name and the area in which you work (after 2011) \newline
P: Please indicate your occupation/professional activity. Provide additional explanations in keywords. \\ \hline

2020-2024 & 
ACS: \textit{No change in question wording} & 
M: For the paper version: Please describe your current occupation in keywords. What is the job title of your current occupation? \newline
M: For the online version: What is the job title of your current occupation? Please describe your current occupation in keywords. \newline
P: Please indicate which occupation/professional activity you performed in the week from May 9 to 15, 2022. Provide additional explanations in keywords. \\
\end{longtblr}
}

\renewcommand{\thetable}{1B}
{\scriptsize
\begin{longtblr}[
  caption={Survey wordings of occupational task questions and whether examples are included. The table summarizes how various German surveys word their open-ended occupation items and whether examples or classification instructions are provided in the question stem or interviewer instructions},  
  label={tab:appendix-1B},  
]{
  colspec={p{2.5cm}|p{4.5cm}|p{4.5cm}|p{1.5cm}},
  rowhead=1
}
\hline
Survey / Instrument & First question wording & Second or third question wording (if applicable) & Examples? \\
\hline
Demographic Standards (2024) \citep{demographische2024standards} & Welche \textbf{berufliche Tätigkeit} üben Sie aus? Wenn Sie nicht mehr Voll- oder Teilzeit erwerbstätig sind: Welche Tätigkeit haben Sie bei Ihrer früheren hauptberuflichen Erwerbstätigkeit zuletzt ausgeübt? &
\textbf{Second question:} Bitte beschreiben Sie diese berufliche Tätigkeit genau. 
\newline \textbf{Third question:} Hat dieser Beruf noch einen besonderen Namen? 1: ja und zwar: / nein & No \\

Mikrozensus (online version) (2023) \citep{bernhard_hochstetter_berufsklassifizierung_nodate} & Welche \textbf{Berufsbezeichnung} hat Ihre gegenwärtige Tätigkeit? \tiny{Z. B.: Modeverkäufer/-in; Grundschullehrer/-in; Reiseverkehrskaufmann/-frau; Bauingenieur/-in; Elektronikmechaniker/-in; Bauhilfsarbeiter/-in; Krankenpfleger/-in} & & Yes, in the stem of the question  \\

Mikrozensus (paper version) (2025) \citep{mikro2025} & 
Bitte beschreiben Sie Ihre gegenwärtige \textbf{Tätigkeit} in Stichworten. \tiny{z.B. Verkauf von Kleidung; Kinder an der Grundschule unterrichten; Kundinnen und Kunden über Reiseangebote beraten und informieren; Bauwerke im Hochbau entwerfen oder planen; Elektronische Schaltungen aufbauen und prüfen; Beton, Gips und Mörtel mischen; Patientinnen und Patienten (vor, während und nach Operationen) betreuen und versorgen} &
\newline \textbf{Second question:} Welche \textbf{Berufsbezeichnung} hat Ihre gegenwärtige Tätigkeit?
\tiny{z.B. Modeverkäufer/-in; Grundschullehrer/-in; Reiseverkehrskaufmann/-frau; Bauingenieur/-in; Elektronikmechaniker/-in; Bauhilfsarbeiter/-in; Krankenpfleger/-in} & Yes, in the stem of the question  \\

ALLBUS 2023 \citep{GESIS_ALLBUS_2023_CAWI}& Welche \textbf{berufliche Tätigkeit} üben Sie in Ihrem Hauptberuf aus? Bitte beschreiben Sie Ihre berufliche Tätigkeit genau.& Hat dieser \textbf{Beruf}, diese \textbf{Tätigkeit} noch einen besonderen Namen? & No \\

NEPS 2024 \citep{FDZLIfBi_NEPS_SC6_W15_2024} & Sagen Sie mir bitte, welche \textbf{berufliche Tätigkeit} Sie da ausgeübt haben! \tiny{According to interviewer instructions an exact job title is required.} & Können Sie mir diese berufliche Tätigkeit noch konkreter benennen? Hat das, was Sie gemacht haben bzw. machen, noch eine genauere Bezeichnung? & No \\

GLES 2025 \citep{GESIS_GLES_RCS_2025} & Welche \textbf{berufliche Tätigkeit} übten Sie in Ihrem Hauptberuf aus? Hat diese Tätigkeit einen besonderen Namen? \newline \tiny{Bitte geben Sie die genaue Tätigkeitsbezeichnung an, also z.B. nicht "kaufmännische/r Angestellte/r", sondern: "Speditionskauffrau/-mann", nicht "Arbeiter/in", sondern: "Maschinenschlosser/in". Wenn Sie Beamter/-in waren, geben Sie bitte Ihre Amtsbezeichnung an, z.B. "Polizeimeister/in" oder "Studienrat/-rätin". Wenn Sie Auszubildende/r waren, geben Sie bitte Ihren Ausbildungsberuf an.} & & Yes, in the instructions \\

SOEP 2023 \citep{SOEP2025} & Welche \textbf{berufliche Tätigkeit} üben Sie derzeit aus? \newline \tiny{Bitte geben Sie die genaue Tätigkeitsbezeichnung an, also z.B. nicht „kaufmännische Angestellte“, sondern: „Speditionskauffrau”, nicht „Arbeiter“, sondern: „Maschinenschlosser“. Wenn Sie Beamtin oder Beamter sind, geben Sie bitte Ihre Amtsbezeichnung an, z.B. „Polizeimeisterin“, oder „Studienrat“. Wenn Sie Auszubildende oder Auszubildender sind, geben Sie bitte Ihren Ausbildungsberuf an.} & & Yes, in the instructions \\

FReDA 2021 \citep{FReDA_W1B_2021}& Welche \textbf{berufliche Tätigkeit} üben Sie derzeit hauptsächlich aus? Wenn Sie derzeit nicht erwerbstätig sind, welche berufliche Tätigkeit haben Sie zuletzt in Ihrer hauptsächlichen Erwerbstätigkeit ausgeübt? \newline \tiny{Bitte geben Sie die genaue Tätigkeitsbezeichnung an, also z.B. nicht „kaufmännische/r Angestellte/r“, sondern „Speditionskaufmann bzw. -frau“, nicht „Arbeiter/in“, sondern „Maschinenschlosser/in“. Wenn Sie Beamter/in sind, geben Sie bitte Ihre Amtsbezeichnung an, z.B. „Polizeimeister/in“ oder „Studienrat/-rätin“. Wenn Sie Auszubildende/r sind, geben Sie bitte Ihren Ausbildungsberuf an.} & Hat dieser Beruf noch eine besondere \textbf{Bezeichnung}?  & Yes, in the instructions \\

GEDA 2019/2020 \citep{RKI_GEDA_2019_2020_EHIS}& Welchen \textbf{Beruf} üben Sie derzeit hauptsächlich aus? \newline \tiny{Geben Sie bitte die genaue Berufsbezeichnung an, nicht den Ausbildungsabschluss oder Rang. Zum Beispiel: - Blumenverkäuferin (nicht Verkäuferin) - Maurer (nicht Bauarbeiter) - Grundschullehrer (nicht Lehrer oder Beamte) - Unternehmensberaterin (nicht Betriebswirtin)} & Um die Einordnung Ihres Berufes zu erleichtern, geben Sie bitte zusätzliche Erläuterungen in
Stichworten an. \tiny{Zum Beispiel: als Blumenverkäuferin: Kundenberatung, Verkauf, Verpacken von Pflanzen; als Zollbeamter: Zollfahndung, Einsatzplanung, Pressearbeit; Falls Sie Führungsaufgaben wahrnehmen, vermerken Sie dies auch.; als KFZ-Mechaniker: Wartung, Instandsetzung, Ausrüstung von Kraftfahrzeugen, Leitung der
Werkstatt} & Yes, in the instructions \\

GESIS Panel (from 2023 on) \citep{GESIS_Panel_ZG_2023} & Welchen \textbf{Beruf} üben Sie zur Zeit aus? \newline \tiny{Bitte geben Sie die genaue Tätigkeitsbezeichnung an. Schreiben Sie z.B. "Speditionskauffrau" und nicht "kaufmännische Angestellte", oder "Maschinenschlosser" statt "Arbeiter". Wenn Sie Beamte/r sind, geben Sie bitte Ihre Amtsbezeichnung an, z.B. "Polizeimeister" oder "Studienrat". Wenn Sie Auszubildende/r sind, geben Sie bitte Ihren Ausbildungsberuf an.}&
Bitte beschreiben Sie noch kurz lhre \textbf{berufliche Tätigkeit} und Ihre wesentlichen Aufgaben.
\tiny{Beispiele sind "Regale mit Produkten auffüllen und Inventur machen", "Patienten versorgen, Medikamente geben, Vitalzeichen überwachen", "Patienten betreuen,
Zähne und Zahnfleisch behandeln", "Überwachung des Warenbestands in der Damenabteilung, Bedienen von Kundinnen und Kassiertätigkeit"}. & Yes, in the instructions \\
\hline
\end{longtblr}
}

\setcounter{figure}{0}  

\renewcommand{\thefigure}{1A}
\begin{figure}[htbp]
\centering
\raisebox{-0.5\height}{\includegraphics[width=\textwidth]{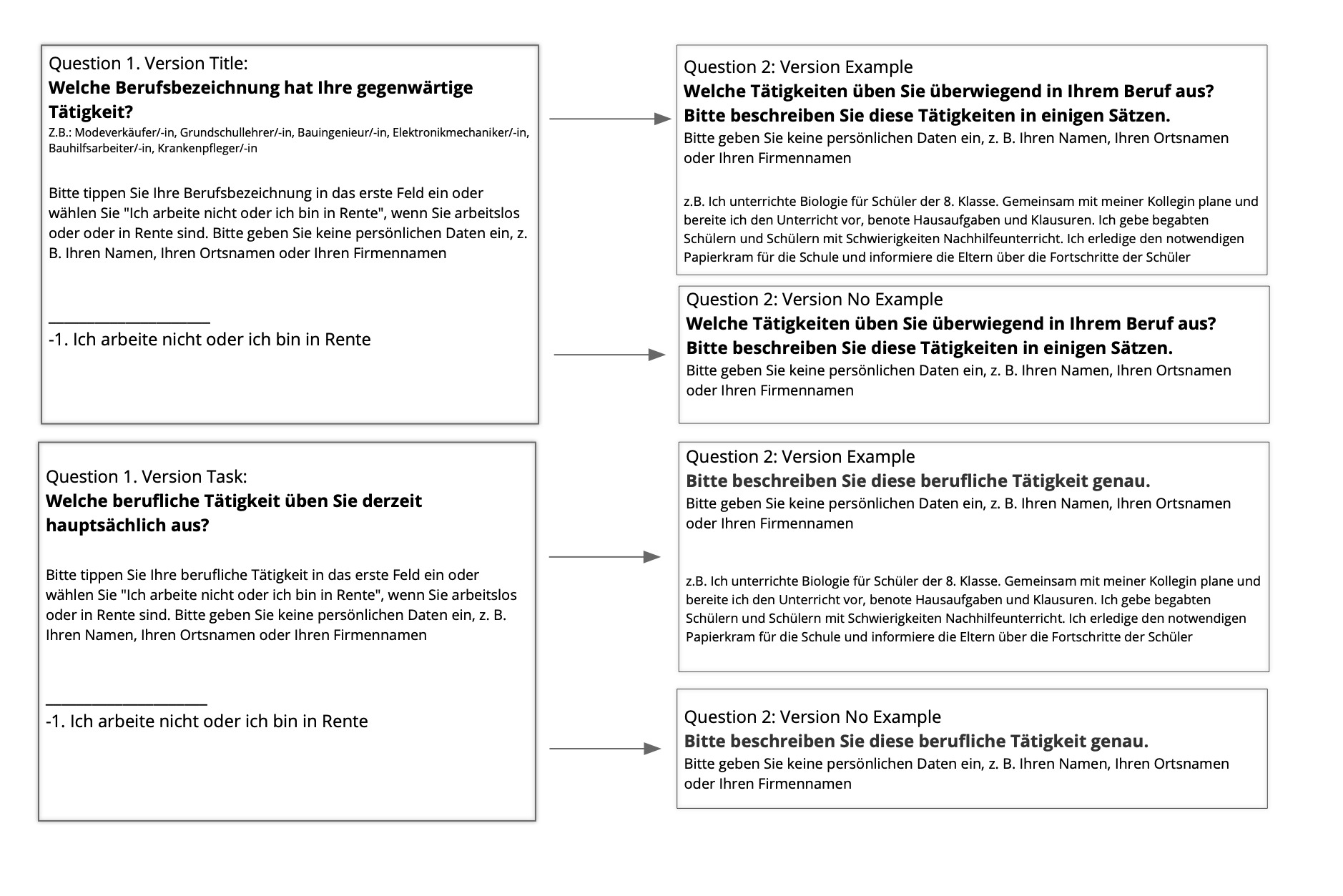}}
\caption{\label{fig:experiment.german}Design of the experiment in German.}
\end{figure}

\renewcommand{\thefigure}{2A}
\begin{figure}[htbp]
\centering
\raisebox{-0.5\height}{\includegraphics[width=0.8\textwidth]{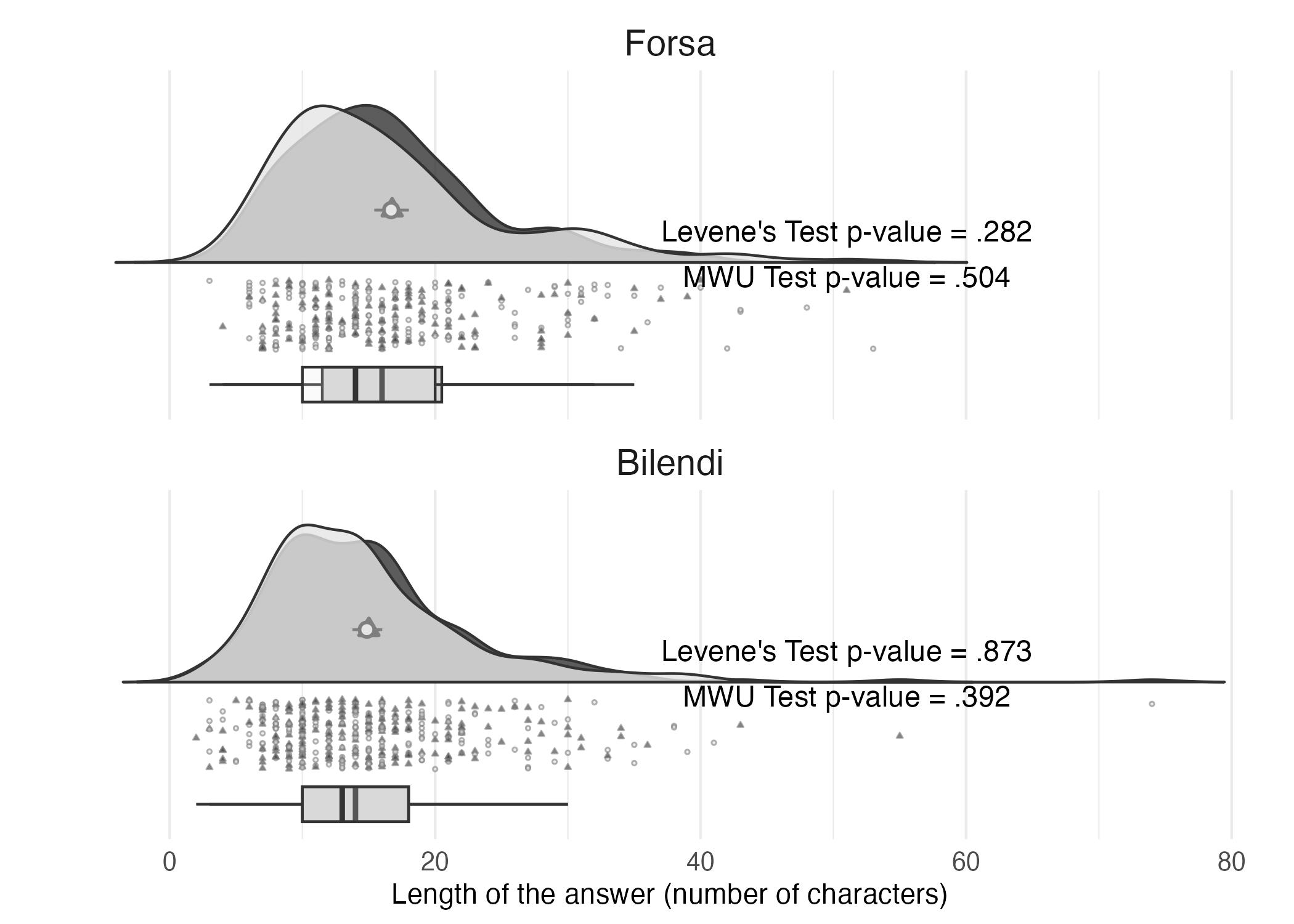}}
\caption{\label{fig:diversity.1} Distribution and characteristics of answers to the Question 1 (job title vs occupational task) based on length (in characters).}
{\scriptsize
\noindent 
\textnormal{
The figure shows the length of answers (in characters) for two Question 1 versions: "Job Title" (dark grey) and "Occupational Task" (light grey). Density curves represent the distribution of answer lengths, with boxplots below summarizing the central tendency and spread. Statistical tests (Levene’s Test and MWU Test) suggest no evidence of differences in variability/median answer length between the two question versions.\\[-0.5em]}
}
\end{figure}

\renewcommand{\thefigure}{3A}
\begin{figure}[htbp]
\centering
\includegraphics[width=0.7\textwidth]{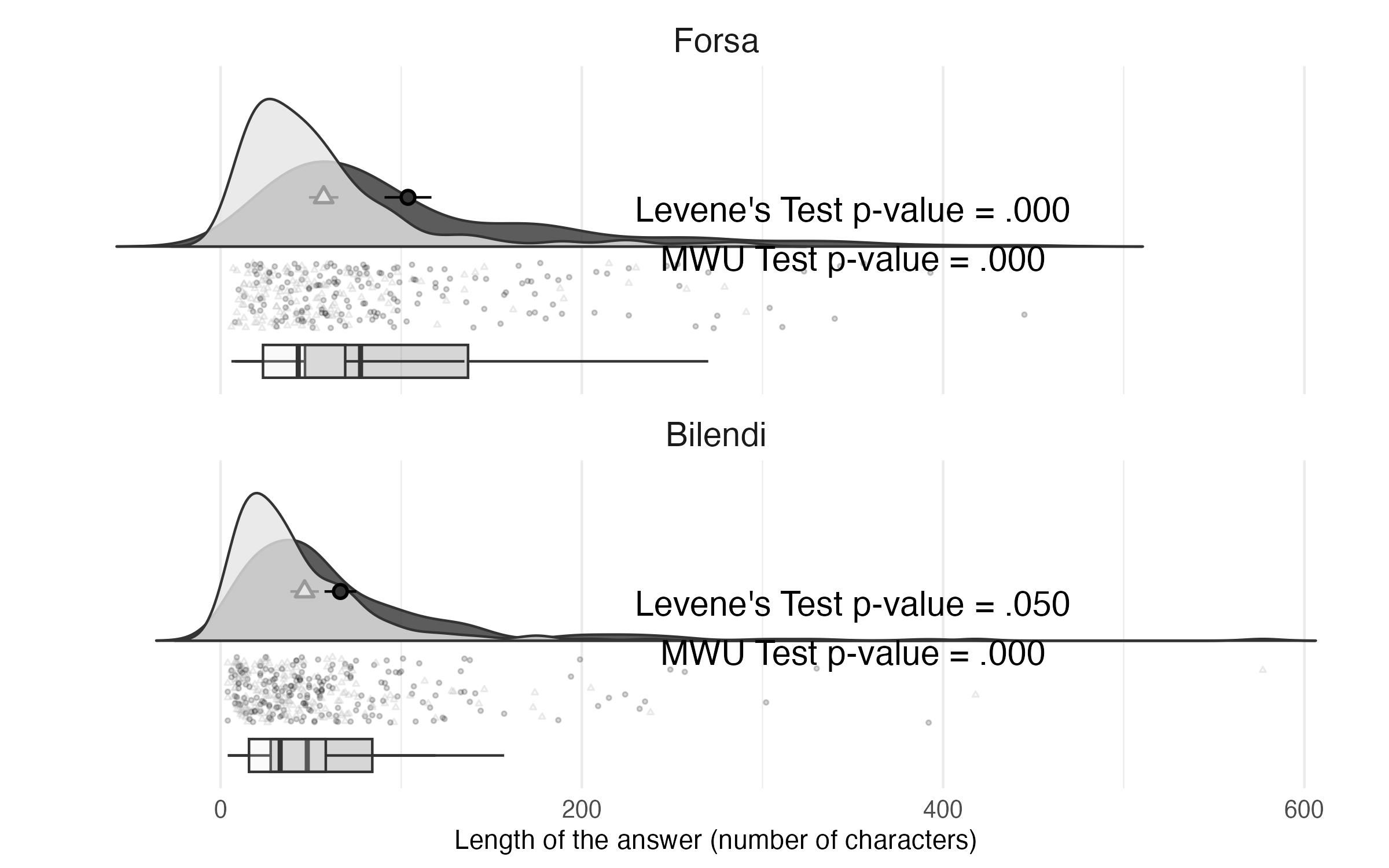}
\caption{\label{fig:length.2} Distribution and characteristics of answers to the Question 2 (example vs no example) based on length (in characters).}
{\scriptsize
\noindent 
\textnormal{
The figure shows the length of answers (in characters) for two Question 2 (Detailed job descriptions) versions: "Example" (green) and "No example" (purple). Density curves represent the distribution of answer lengths, with boxplots below summarizing the central tendency and spread. Results from statistical tests (Levene’s Test and Mann-Whitney U Test) indicate evidence of differences in variability and/or median answer lengths between the two question versions.\\[-0.5em]}
}
\end{figure}

\renewcommand{\thefigure}{4A}
\begin{figure}[htbp]
\centering
\raisebox{-0.5\height}{\includegraphics[width=\textwidth]{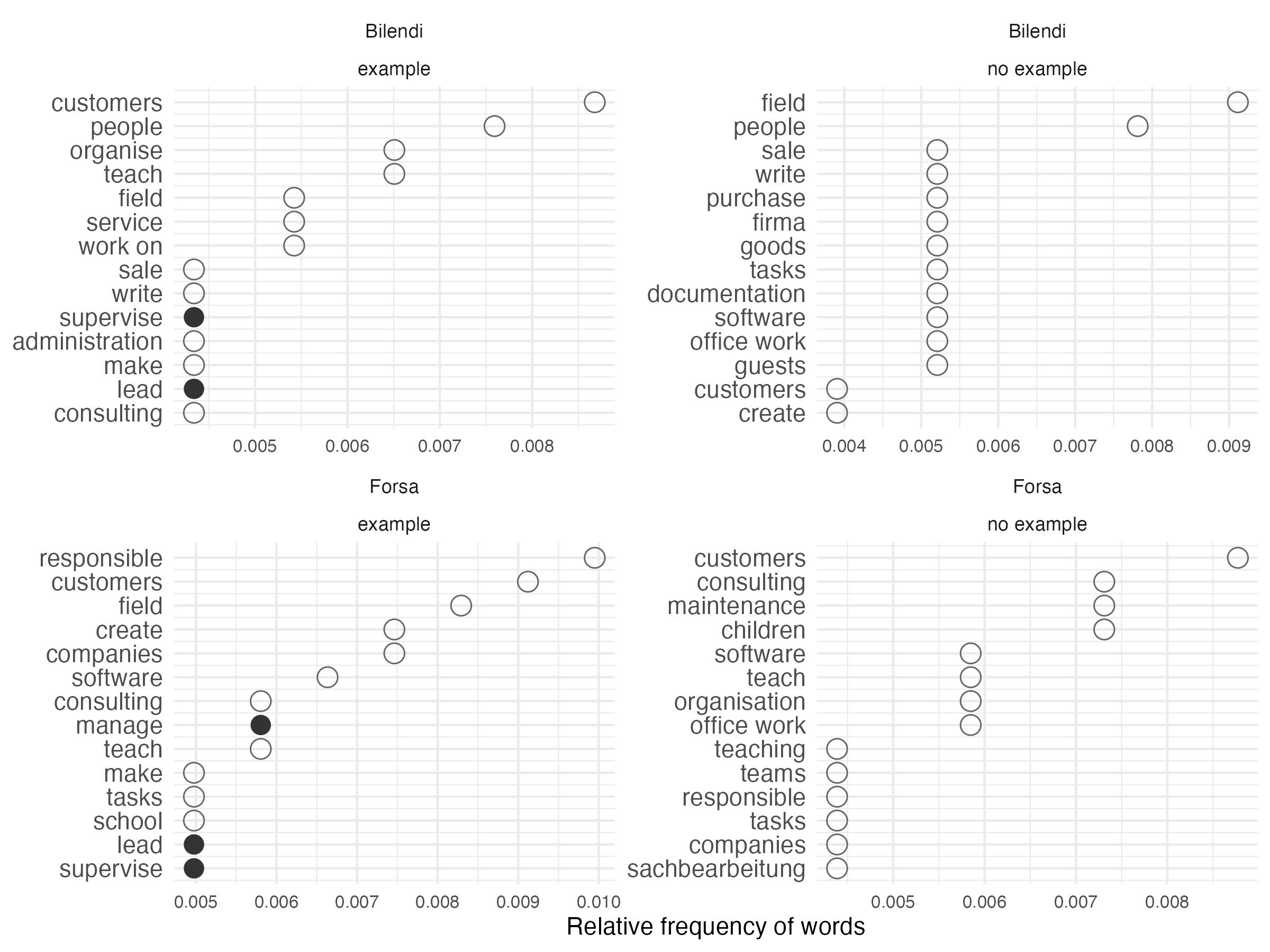}}
\caption{\label{fig:word.frequency}Comparative Analysis of Term Proportions in Responses With and Without Examples.}
\end{figure}

\renewcommand{\thefigure}{\arabic{figure}}
\setcounter{figure}{0}

\end{document}